\documentstyle[12pt]{article}
\newlength{\extraspace}
\newlength{\extraspaces}
\newcommand{\be}{\begin{equation}}
\newcommand{\ee}{\end{equation}}
\newcommand{\br}{\begin{eqnarray}}
\newcommand{\er}{\end{eqnarray}}
\newcommand{\ba}{\begin{array}}
\newcommand{\ea}{\end{array}}
\newcommand{\bi}{\begin{itemize}}
\newcommand{\ei}{\end{itemize}}
\newcommand{\bn}{\begin{enumerate}}
\newcommand{\en}{\end{enumerate}}
\newcommand{\ul}{\underline}
\newcommand{\ol}{\overline}

\begin{document}
\pagestyle{empty}
\begin{titlepage}
\begin{flushright}
NSF-ITP-94-68\\
UTPT-94-18\\
hep-ph/9407244
\end{flushright}
\vspace{2.5cm}
\begin{center}
{\LARGE A dynamical origin for the top mass}\\ \vspace{40pt}
{\large B. Holdom\footnote{holdom@utcc.utoronto.ca}}
\vspace{0.5cm}

{\it Department of Physics\\ University of Toronto\\
Toronto, Ontario\\Canada M5S 1A7}
\vspace{0.5cm}

{\it Institute for Theoretical Physics\\ University of California\\
Santa Barbara, California 93106-4030}
\vskip 2.1cm
\rm
\vspace{25pt}
{\bf ABSTRACT}

\vspace{12pt}
\baselineskip=18pt
\begin{minipage}{5in}

We describe a dynamical scheme by which isospin breaking feeds strongly into
the $t$ and $b$ masses and not into the $W$ and $Z$ masses.  The third family
feels a new gauge interaction broken close to a TeV.

\end{minipage}
\end{center}
\vfill
\end{titlepage}
\pagebreak
\baselineskip=18pt
\pagestyle{plain}
\setcounter{page}{1}

A large top quark mass presents an interesting puzzle to those who think that
the origin of quark and lepton masses lies in gauge theory dynamics.  How
does the dynamics responsible for the isospin breaking in the $t$-$b$ masses
not also feed into the $W$ and $Z$ masses and contribute substantially to
$\Delta \rho $?  A typical approach to isospin breaking is to build it
explicitly into the theory and have the right-handed $t$ and $b$ fields feel
intrinsically different dynamics in the underlying theory.  We will proceed in
the opposite direction and contemplate a more dynamical origin of isospin
breaking in the hope that this leads to more economical and appealing
theories.  In the technicolor context it was noticed that the dynamical
techniquark masses can feel the effect of isospin breaking in physics at
higher scales via effective 4-techniquark operators.  This could lead to
isospin breaking in techniquark condensates and thus quark and lepton
masses.  The trouble is that some fine tuning in the strength of the
4-techniquark operators is necessary to prevent this isospin breaking from
contributing to
$\Delta\rho $.  This fine tuning becomes quite unbearable for a large top
mass.\cite{A}

In this note we will consider a mechanism which requires that the third
family feel a new gauge interaction which is broken at relatively low scales,
close to a TeV.  In the picture we actually develop this breaking will
coincide with electroweak symmetry breaking.  We call this new interaction
hypercolor (HC), and we will assume that it has a slowly walking gauge
coupling.\cite{K}  The basic idea is that isospin breaking originates
dynamically on energy scales of order 100 TeV and manifests itself in
effective 4-hyperquark operators.  One isospin breaking operator in
particular remains relevant at lower energy scales due to the walking nature
of HC.  HC eventually breaks via a condensate which also breaks electroweak
symmetry, and members of the third family become gauge singlets under the
unbroken subgroup of HC.  The 4-hyperquark operator, in combination with the
condensate, then directly feeds a mass to the $t$ and not the $b$.  We will
argue that because of the particular structure of the 4-hyperquark operator,
neither it or other operators it induces will contribute significantly to
isospin breaking in the electroweak breaking condensate.

The fermion fields participating in the electroweak breaking condensate will
also be singlets under the unbroken subgroup of HC.  In the end these fields
are nothing more than an ordinary fourth family of fermions.  The fourth
family quarks, $t'$ and $b'$ will have masses of order 1 TeV and they will be
approximately degenerate.  Our goal is to show how this can occur in a
dynamical context while at the same time having large isospin breaking in the
third family quark masses.

We will take $SU(3{)}_{H}$ as the HC gauge group.  There are two families of
hyperfermions, each with hyperquarks and hyperleptons transforming under
$SU(3)_C\times SU(2)_L\times U(1)_Y$ in the standard way.  One family of
hyperfermions denoted by $H$ all transform as a {\bf 3} under HC while the
other family, $\ul{H}$, all transform as
$\bf \ol{3}$.  The hyperquarks in these two families are $Q = (U,D)$ and
$\ul{Q} = (\ul{U},\ul{D})$.  The following condensates are assumed to occur
on the TeV scale,
\be\langle{\ol{\ul{U}}}_{L}^{A}{U}_{R}^{B}\rangle\approx
\langle{\ol{\ul{D}}}_{L}^{A}{D}_{R}^{B}\rangle\propto{\delta }^{A3}{\delta
}^{B3}\label{c}\ee where $A$ and $B$ are HC indices.\footnote{Hermitian
conjugate terms are implicitly assumed throughout this paper.}  This breaks
$SU(2)_L\times U(1)_Y$ in the standard fashion.  The condensate lies in the
$\bf 6$ of $SU(3{)}_{H}$ which thus breaks; we will call the unbroken
subgroup $SU(2{)}_{M}$ or metacolor.  We will discuss the question of why
this HC breaking condensate forms later.  Whatever this dynamics is, it
should preserve isospin to good approximation.

The HC triplets contain the fields $({H}^{1},{H}^{2},{H}^{3})\equiv
({M}^{1},{M}^{2},f)$ and
\linebreak $({\ul{H}}^{1},{\ul{H}}^{2},{\ul{H}}^{3})\equiv
({\ul{M}}^{1},{\ul{M}}^{2},\ul{f})$.  ${M}^{a}$ and ${\ul{M}}^{a}$ are two
metacolored families and $f$ and $\ul f$ denote two families which are
metacolor singlets.  The latter contain the quark doublets $q$ and
$\ul{q}$.  From the condensate we see that the Dirac spinors to be associated
with the massive fourth family quarks are of the form
$[{\ul{q}}_{L},{q}_{R}]$.  From now on we denote these fields by
$q'=(t',b')$.  Dirac spinors of the form $[{q}_{L},{\ul{q}}_{R}]$ will be
relabeled as the third family quarks $q=(t,b)$.  Note that the two light
families are HC singlets and are outside the present discussion.

Most of our discussion will also not involve the lepton sector, since the
mechanism we are describing for the quarks need not also be occurring for the
leptons.  We will assume that the fourth family leptons are somewhat lighter
than the fourth family quarks.  Then the dynamical $t'$ and $b'$ masses make
the main contribution to the $W$ and $Z$ masses and the associated decay
constant is $F\approx$ 145 GeV.  From this we may estimate the $t'$ and
$b'$ masses,
\be {m}_{q'}\approx \sqrt {3}F{\frac{{m}_{\rho }}{2{f}_{\pi }}}\approx 1
\;{\rm TeV}.\ee  The $\sqrt 3$ accounts for ${m}_{q'}$ scaling as
${N}_{c}^{-1/2}$ when $F$ is held fixed; in QCD $N_c=3$ whereas here $N_c$ is
effectively 1.

Now consider the presence of the isospin-violating (IV) but $SU(3)_H\times
SU(3)_C\times SU(2)_L\times U(1)_Y$ preserving operator,
\be{\frac{1}{{\Lambda
}_{IV}^{2}}}\epsilon_{ij}{\ol{Q}}_{L}^{Ai}{D}_{R}^{A}{\ol{\ul{Q}}}_{L}^{Bj}
{\ul{U}}_{R}^{B}.\label{a}\ee
$SU(2)_L$ and HC indices are shown.  This operator originates in ``extended
HC" (EHC) physics on energy scales of order 100 TeV. It cannot have a
perturbative origin; it must arise dynamically as we discuss more below.
This dynamics breaks $SU(2)_R$ in a maximal way by not generating the related
operator
$\epsilon_{ij}{\ol{Q}}_{L}^{Ai}{U}_{R}^{A}{\ol{\ul{Q}}}_{L}^{Bj}{\ul{D}}_{R}^{B}$.
Although operator (\ref{a}) is generated at the EHC scale, we will take the
actual value of ${\Lambda }_{IV}$ to correspond to renormalization of the
operator at the HC breaking scale.

The operator (\ref{a}) contains the term
\be{\frac{1}{{\Lambda
}_{IV}^{2}}}\epsilon_{ij}{\ol{q}}^i_{L}{b'}_{R}{\ol{q'}}^j_{L}{t}_{R}.\ee  In
combination with the $q'$ condensate (\ref{c}) this produces a $t$ mass, but
no $b$ mass.  When the color indices appear in the appropriate way the result
is
\be{m}_{t}\approx {\frac{3\langle\ol{b'}b'\rangle}{4{\Lambda }_{IV}^2}}.\ee
(There is no color sum inside the condensate.)  The condensate is
renormalized at the HC scale, and thus we estimate
\be\langle\ol{b'}b'\rangle \approx 4\pi \kappa \sqrt {3}{F}^{3}.\ee  $\kappa$
is a factor allowed to vary between 1 and 2, corresponding to a correction
factor of 1.7 in QCD.\cite{B}  The $\sqrt {3}$ accounts for the condensate
scaling as ${N}_{c}^{-1/2}$ when $F$ is held fixed (see above and \cite{J}).
$m_t=175$ GeV would imply that \be{\Lambda }_{IV}\approx  (550-750)\;{\rm
GeV}.\ee

We see then that the coefficient of the 4-hyperfermion operator (\ref{a}) is
significantly smaller than a typical coefficient of a 4-hyperfermion operator
produced by HC dynamics.  By ``naive dimensional analysis",\cite{C,J} such a
coefficient is of order $1/F^2$.

On the other hand this value for ${\Lambda }_{IV}$ is consistent with this
operator originating at the higher EHC scale if HC is a walking theory.  HC
interactions will generate an anomalous dimension of order
$2{\gamma }_{m}$ where ${\gamma }_{m}$ is the anomalous dimension of either of
the HC singlet mass operators ${\ol{Q}}_{L}^{A}{D}_{R}^{A}$,
${\ol{\ul{Q}}}_{L}^{A}{\ul{U}}_{R}^{A}$.  In a walking theory ${\gamma }_{m}$
is close to unity.\cite{K}  Indeed at one loop in Landau gauge, operator
(\ref{a}) does not mix with other operators and its anomalous dimension is
exactly equal to $2{\gamma }_{m}$.  Thus we expect that operator (\ref{a}) is
strongly enhanced and it may be a relevant operator (or very nearly so) in
the effective theory well below the EHC scale.

The crucial question is whether the presence of this operator implies that
there are other significant contributions to $\Delta\rho$ besides the one
coming from the $t$ mass.  In fact operator (\ref{a}) directly contributes to
another $SU(2)_R$ violating operator:
\be{\frac{1}{{\Lambda
}_{IV}^{'2}}}{\ol{\ul{Q}}}_{L}^{A}{D}_{R}^{B}{\ol{D}}_{R}^{B}{\ul{Q}}_{L}^{A}\label{b}.\ee
This is generated from a one loop diagram with operator (\ref{a}) and its
hermitian conjugate at two vertices.  If we repeat the procedure by putting
operator (\ref{b}) and its conjugate in a loop, then we can also generate the
operator: \be{\frac{1}{{\Lambda }_{IV}^{''2}}}{\left[{\ol{D}}_{R}{\gamma
}_{\mu }{D}_{R}\right]}^2\label{z}.\ee  We again take the values of ${\Lambda
}_{IV}^{'}$ and ${\Lambda }_{IV}^{''}$ to correspond to renormalization at
the HC scale.  Operator (\ref{b}) directly contributes to the gap equation
for the condensate (\ref{c}), resulting in
$b'$-$t'$ mass splitting.  The latter is of order
$\langle \ol{b'}b'\rangle /{\Lambda }_{IV}^{'2}$, and $\Delta\rho$ is
proportional to the square of this mass splitting in the standard way.
Operator  (\ref{z}) directly induces the
$\Delta\rho$ term in the chiral Lagrangian,
$F^2 {\left[{\rm Tr}\{{U}^{\dagger }{D}_{\mu }U{\tau }_{3}\}\right]}^2$.  This
contribution to $\Delta\rho$ is of order ${F}^{2}/{\Lambda }_{IV}^{''2}$.

We have just said that the minimal contribution to the operators (\ref{b}) and
(\ref{z}) implied by the existence of operator (\ref{a}) occurs at the one and
three loop level respectively.  When the loop momenta in these diagrams is of
order the EHC scale, the resulting effective operators must then be run down
to the HC scale.  But we see that the one-loop renormalization of these
operators is completely different than operator (\ref{a}).  Besides causing
mixing with other operators, the one-loop renormalization does not strongly
enhance operators (\ref{b}) and (\ref{z}).  This suggests that even beyond
one loop, the enhancement of operators (\ref{b}) and (\ref{z}) can be
nonexistent or at least much less than the enhancement of operator
(\ref{a}).  Considering also that the operators (\ref{b}) and (\ref{z}) may
start off smaller at the EHC scale, it seems very plausible that ${\Lambda
}_{IV}^{'2}$ and
${\Lambda }_{IV}^{''2}$ are sufficiently large to cause little contribution to
$\Delta \rho $.

If we again consider the one loop diagram which generates the operator
(\ref{b}), we find also a long distance contribution when the loop momentum
is of order the HC scale.  Here naive dimensional analysis\cite{C} can be
used, which implies that the long distance contribution to $1/{\Lambda
}_{IV}^{'2}$ is of order
${F}^{2}/{\Lambda }_{IV}^{4}$.  The resulting contribution to $b'$-$t'$ mass
splitting can then be written roughly as $({{F}^{2}}/{{\Lambda
}_{IV}^{2}}){m}_{t}$.  The important point is that it is small compared to the
$t$-$b$ mass splitting, as is the corresponding contribution to $\Delta \rho
$.  Similarly, naive dimensional analysis implies that the long distance
contribution to $1/{\Lambda }_{IV}^{''2}$ is of order
${F}^{6}/{\Lambda }_{IV}^{8}$ and is thus negligible.

The conclusion is that it appears natural for the largest contribution to
$\Delta \rho $ to be the one associated with the top quark mass.

In order to have some idea where the EHC operators come from we indicate a
possible gauge group and particle content above the EHC scale.\footnote{The
Pati-Salam gauge group, ``422", may actually be replaced by some subgroup
which contains ``321".}\cite{D,E}
\be U(1{)}_{A}\times U(4{)}_{S}\times SU(4{)}_{PS}\times SU(2{)}_{L}\times
SU(2{)}_{R}\ee
\be\ba{c}(+1,{\bf 4},{\bf 4},{\bf 2},{\bf 1}{)}_{L}\\ (-1,{\bf 4},{\bf
4},{\bf 1},{\bf 2}{)}_{R}\\ (-1,{\bf\ol{4}},{\bf 4},{\bf 2},{\bf 1}{)}_{L}\\
(+1,{\bf\ol{4}},{\bf 4},{\bf 1},{\bf 2}{)}_{R}\ea\label{y}\ee The ``sideways"
$U(4)_{S}$ includes a $U(1)$ with opposite vector charges for fermions in the
${\bf 4}$ and ${\bf\ol{4}}$.  The important point is that we have a chiral
gauge theory; any bilinear condensate breaks some gauge symmetry, even in the
absence of the weak
$SU(2{)}_{L}\times SU(2{)}_{R}$.   Because this is not a vector-like theory
we avoid the Vafa-Witten result,\cite{F} and can thus contemplate breaking
isospin via the dynamical breaking of $SU(2{)}_{R}$.  We take one
manifestation of this to be EHC scale Majorana masses for the right-handed
neutrinos and hyperneutrinos, which leaves unbroken the correct hypercharge
$U(1)_Y$.  These condensates will be the only bilinear ones allowed by the
unbroken symmetries.

We take another manifestation to be the $SU(2{)}_{R}$ violating condensate
\be\langle\epsilon_{ij}{\ol{Q}}_{L}^{Ai}{D}_{R}^{A}{\ol{\ul{Q}}}_{L}^{Bj}
{\ul{U}}_{R}^{B}\rangle.\label{d}\ee  This signals the dynamical generation of
a 4-point function which otherwise vanishes in perturbation theory.  For
external momenta small compared to the EHC scale, the momentum dependence of
the 4-point function should be well described by the effective theory in
which the corresponding local operator times a coefficient of order
$1/{\Lambda }_{EHC}^{2}$ appears.  The result at HC scales is the desired
effective operator (\ref{a}).

We argue that the appearance of condensate (\ref{d}) can be considered
natural (in the sense of no fine-tuning), and we may illustrate this point by
a toy scalar field potential.  Consider a scalar field ${\phi
}_{ik\ul{i}\ul{k}}$ where $i$ and
$\ul{i}$ are $SU(2)_L$ doublet indices and $k$ and $\ul{k}$ are
$SU(2)_R$ doublet indices.  We then construct the most general
$SU(2{)}_{L}\times SU(2{)}_{R}$ invariant, quadratic plus quartic scalar
potential.  An underlined index cannot be contracted with an index not
underlined; this reflects the two flavors of hyperquark doublets in the
model.  For a range of values for the parameters we find that at the minimum
of the potential $SU(2{)}_{R}$ breaks but not $SU(2{)}_{L}$.  The vacuum
expectation value could be\footnote{Weak vacuum alignment may help choose the
direction of the vev.}
\be\langle {\phi }_{ik
\ul{i}\ul{k}}\rangle \propto {\left({\begin{array}{cc}0&-1\\
1&0\end{array}}\right)}_{i\ul{i}}{\left({\begin{array}{cc}0&0\\
1&0\end{array}}\right)}_{k\ul{k}},\ee which corresponds to our desired
condensate (\ref{d}).

Below the EHC scale we take the unbroken gauge symmetry to be\footnote{The
$U(1)_A$ must be broken since it couples to light fermions; this can be
accomplished by the EHC condensate
$\langle{\ol{Q}}_{L}^{A}{Q}_{R}^{A}{\ol{\ul{Q}}}_{R}^{B}
{\ul{Q}}_{L}^{B}\rangle$.}
\be U(3{)}_{H}\times SU(3)_C\times SU(2{)}_{L}\times U(1{)}_{Y}.\ee Operators
of interest for the 1 TeV HC breaking and which can be produced
perturbatively at the EHC scale are
\be{\ol{Q}}_{L}^{A}{Q}_{R}^{A}{\ol{{Q}}}_{R}^{B} {{Q}}_{L}^{B}\;\;\;{\rm
and}\;\;\;{\ol{\ul Q}}_{L}^{A}{\ul Q}_{R}^{A}{\ol{\ul{Q}}}_{R}^{B}
{\ul{Q}}_{L}^{B}.\label{e}\ee  Like operator (\ref{a}) they are strongly
enhanced as they are run down to the HC scale, and can thus play a role in
the HC symmetry breaking.  If the contribution from the massive $U(1)_A$
boson dominates then these operators are produced with the desired signs to
resist the formation of the HC singlet condensates
$\langle{\ol{Q}}_{L}^{A}{Q}_{R}^{A}\rangle$ and $\langle{\ol{\ul
Q}}_{L}^{A}{\ul Q}_{R}^{A}\rangle$.  It is this $U(1)_A$ effect which can
drive the HC dynamics away from what would naively be the most attractive
channel; we assume that the HC dynamics instead produces the HC breaking
condensate (\ref{c}).  This latter condensate leaves unbroken the gauge
symmetry  \be U(2{)}_{M}\times SU(3{)}_{C}\times U(1{)}_{EM}.\ee

We have been describing possible dynamics responsible for the mass of a fourth
family and the top quark.  Is it at least conceivable that all the other quark
and lepton masses can be generated within this same scheme?  One possibility
has been described in \cite{E}.  We would like to briefly describe an
alternative way to generate light fermion masses involving slightly less
breaking of the original gauge symmetries.  The result is the massless
metaphoton in $U(2)_M$ which would have interesting implications of its
own.\cite{G}  We relegate this discussion to an Appendix since it is quite
independent of the main point of this paper.

The main point has been to develop a dynamical scheme by which a large top
mass is consistent with a small $\Delta\rho$.  The scheme appears to be both
natural and economical.  Of course it is impossible to claim that the
strongly interacting chiral gauge theory actually behaves the way we have
described.  But the scheme does present a testable picture of the physics in
the 0.1 to 1 TeV range.  Besides the fourth family there is the metacolor
sector,\cite{E} which produces a mass spectrum of bound states lying in this
energy range.  There is also a massive hypercolor gauge boson (the diagonal
generator in $U(3)_H/U(2)_M$) which couples only to the third and fourth
families.

\newpage
\noindent{\Large \bf Appendix}

As described in \cite{E} we will assume that no dynamical metafermion mass
forms, which is guaranteed if certain discrete symmetries remain unbroken.  In
this situation we expect little contribution to the electroweak correction
parameter $S$ from the metacolor sector. 4-metafermion condensates are
sufficient to produce the $b$ and $\tau$ masses, and the $e$, $\mu$, and
$\tau$ neutrinos are naturally light.  The new feature we are considering is
the metaphoton which has opposite vector charges for the two metacolored
families.  A massless metaphoton did not exist in \cite{E} since in that case
the generation of masses for the light two families involved operators
generated at the EHC scale which broke the original $U(1)$ in $U(4)_S$.  We
must therefore find another way to generate masses for the two light families.

We will denote members of the two light families by $\chi$ and $\ul{\chi}$
(which lie in the ${\bf 4}$ and $\bf\ol{4}$ in (\ref{y})).  We first note
that their masses depend on HC breaking effects.  To the extent that EHC
interactions are HC conserving then operators of the form
${\ol{H}}_{L}{H}_{R}{\ol{\chi}}_{R}{\chi}_{L}$ and
${\ol{\ul{H}}}_{L}{\ul{H}}_{R}{\ol{\ul{\chi}}}_{R}{\ul{\chi}}_{L}$ are
generated by broken $SU(4)_S$ gauge boson exchange.  But these are useless for
feeding down mass from the third or fourth family quarks to the first two
families since the $t'$ and $b'$ masses are of the form
${\ol{\ul{H}}}_{L}^{3}{H}_{R}^{3}$ while the $t$ and $b$ masses are of the
form ${\ol{H}}_{L}^{3}{\ul{H}}_{R}^{3}$.

But the question is whether EHC physics is completely immune to the HC
breaking physics.  The walking nature of the HC interaction suggests that
perhaps it is not.  We will entertain the notion that HC breaking should be
reflected in the effective theory all the way up to EHC scales, with the only
constraint being that the HC breaking effects be consistent with HC gauge
boson masses less than a TeV.

We consider the existence on EHC scales of the HC violating operators
\be{\ol{H}}_{L}^3{\ul{H}}_{R}^3{\ol{\ul{\chi}}}_{R}{\chi}_{L}\;\;\;{\rm
and}\;\;\; {\ol{\ul{H}}}_{L}^3{H}_{R}^3{\ol{\chi}}_{R}{\ul{\chi}}_{L}.\ee
These operators will be produced for example if there is a mass splitting
between a certain massive $SU(4)_S$ gauge boson in the $SO(4)$ of $SU(4)_S$
and an associated gauge boson not in the $SO(4)$.  In any case the size of
the coefficients of the operators is of order $1/{\Lambda }_{EHC}^{2}$ times
some suppression factor.  The suppression factor, perhaps of order 0.1 or
0.01, ensures that the feedback into the massive HC gauge bosons is
sufficiently small.

These operators will feed mass down to the two light families.  Note that the
Dirac spinors of one light family have the form $[{\chi}_{L},{\ul{\chi}}_{R}]$
and the other light family, $[{\ul{\chi}}_{L},{\chi}_{R}]$.  As of yet there
is no mass mixing to produce weak mixing angles.  (Nor are there flavor
changing neutral currents.)  Mass mixing between the two heavy families and
the two light families can be induced by other HC violating operators, for
example
${\ol{H}}_{L}^{3}{\ul{H}}_{R}^{3}{\ol{\ul{H}}}_{R}^{3}{\chi }_{L}$.  Such
effects cannot be large, which is consistent with the observed smallness of
the KM mixing involving the third family.  Finally we note that the origin of
Cabibbo mixing between the first two families may be associated with the large
right-handed neutrino masses, ${\nu }_{eR}{\nu }_{eR}$, ${\nu }_{\mu R}{\nu
}_{\mu R}$, ${\nu }_{eR}{\nu }_{\mu R}$.  For example these masses could
communicate with the quark sector via $SU(4)_{PS}$ interactions, thus
producing $u$-$c$ mass mixing.  $d$-$s$ mass mixing and thus
${K}^{o}{\ol{K}}^{o}$ mixing is naturally suppressed.\cite{I}

We are feeding down masses to the light fermions via operators generated at
the EHC scale.  Each light mass will thus be determined by the mass of the
appropriate fermion of the third or fourth family evaluated at the EHC
scale.  Notice that the various third and fourth family fermions are
receiving masses in quite different ways, thus leading to quite different
masses at the EHC scale.  For example, it is interesting to speculate that the
fourth family masses are actually much softer at high energies than the third
family masses.  In this case the first and second families could receive
their masses from the fourth and third families, respectively.  This would
produce more isospin breaking in the second family than in the first, as
observed.

\vspace{3ex}
\noindent {\Large\bf Acknowledgments}
\vspace{1ex}

I thank the organizers of the Weak Interactions Workshop and the Institute for
Theoretical Physics for their hospitality.  I thank Lisa Randall, John
Terning and George Triantaphyllou for discussions.  This research was
supported in part by the National Science Foundation under Grant No.
PHY89-04035 and by the Natural Sciences and Engineering Research Council of
Canada.

\end{document}